\documentclass[sigconf, author]{acmart}

\AtBeginDocument{%
  }
    
\usepackage{comment}
\usepackage{tabularx}
\usepackage{array}
\newcolumntype{Y}{>{\centering\arraybackslash}X}
\usepackage{amsmath}
\usepackage{graphics}
\usepackage{color}
\usepackage{fontawesome}
\usepackage{todonotes}
\usepackage{xspace}
\usepackage{soul}
\usepackage{siunitx}
\usepackage[flushleft]{threeparttable}
\usepackage[title]{appendix}
\usepackage{enumitem}
\usepackage{balance}
\usepackage{xspace}
\usepackage{xcolor}
\usepackage{color, colortbl, soul}
\usepackage{tikz}
\usepackage{caption,subcaption}
\usepackage{fixltx2e}
\usepackage{CJKutf8}
\usepackage{pifont}
\usepackage{listings}
\usepackage[most]{tcolorbox}
\def\ie{\textit{i.e.,}\xspace}

\def\eg{\textit{e.g.,}\xspace}

\emergencystretch=\maxdimen
\setcopyright{none}
\renewcommand\footnotetextcopyrightpermission[1]{}
\author{Menghe Zhang}
\email{mez071@ucsd.edu}
\orcid{0000-0001-8681-4889}
\affiliation{%
  \department{Computer Science and Engineering}
  \institution{University of California San Diego}
  \city{La Jolla}
  \state{CA}
  \country{United States}
}

\author{Chen Chen}
\email{chenchen@ucsd.edu}
\orcid{0000-0001-7179-0861}
\affiliation{%
  \department{Computer Science and Engineering}
  \institution{University of California San Diego}
  \city{La Jolla}
  \state{CA}
  \country{United States}
}

\author{Matin Yarmand}
\email{myarmand@ucsd.edu}
\orcid{0000-0002-2353-3462}
\affiliation{%
  \department{Computer Science and Engineering}
  \institution{University of California San Diego}
  \city{La Jolla}
  \state{CA}
  \country{United States}
}

\author{Anish Rajeshkumar}
\email{arajeshkumar@ucsd.edu}
\orcid{0009-0009-5040-5253}
\affiliation{%
  \department{Mathematics}
  \institution{University of California San Diego}
  \city{La Jolla}
  \state{CA}
  \country{United States}
}

\author{Nadir Weibel}
\email{weibel@ucsd.edu}
\orcid{0000-0002-3457-4227}
\affiliation{%
  \department{Computer Science and Engineering}
  \institution{University of California San Diego}
  \city{La Jolla}
  \state{CA}
  \country{United States}
}

\begin{document}

\title[AcuVR]{AcuVR: Enhancing Acupuncture Training Workflow with Virtual Reality}

\begin{abstract}
Acupuncture is a widely adopted medical practice that involves inserting thin needles into specific points on the body to alleviate pain and treat various health conditions. Current learning practices heavily rely on 2D atlases and practice on peers, which are notably less intuitive and pose risks, particularly in sensitive areas such as the eyes. To address these challenges, we introduce AcuVR, a Virtual Reality (VR) based system designed to add a layer of interactivity and realism. This innovation aims to reduce the risks associated with practicing acupuncture techniques while offering more effective learning strategies. Furthermore, AcuVR incorporates medical imaging and standardized anatomy models, enabling the simulation of customized acupuncture scenarios. This feature represents a significant advancement beyond the limitations of conventional resources such as atlases and textbooks, facilitating a more immersive and personalized learning experience. The evaluation study with eight acupuncture students and practitioners revealed high participant satisfaction and pointed to the effectiveness and potential of AcuVR as a valuable addition to acupuncture training. 
\end{abstract}



\maketitle

\pagestyle{plain}

\section{Introduction}
\label{sec:intro}
Acupuncture is a widely adopted therapeutic practice, especially for providing pain relief through precise insertion of thin needles into specific areas of the body~\cite{kaptchuk2002acupuncture}. It has long been used in clinical practices around the globe, widely employed in over 183 countries and regions worldwide~\cite{world2019global} for 461 symptoms and 972 diseases~\cite{yan2022review}. However, there is a lack of efficient approach to support today's acupuncture training workflow.
When learning acupuncture theories, the heavy reliance on textbooks and atlases can hardly provide novices with an intuitive mental understanding of the 3D anatomical structures.
During learning acupuncture skills, while needling on peer students is a common approach~\cite{janz2011acupuncture, lee2015evaluation, zhang2023vr}, failure to see through the below-surface bodily structure can reduce learning efficacy, and potentially harm patients~\cite{xu2023adverse}.
Given the rising demand for acupuncture services, it is therefore important to design a safe and effective approach to enhance acupuncture training workflow.

Virtual Reality (VR) is promising in healthcare and health science education by enabling more efficient and intuitive interactions across diverse stereoscopic contexts. Moreover, it offers a secure training platform as a substitute for real-life practice on human bodies.
Much prior research has explored how to integrate such affordances into different healthcare and health science education workflows, like contouring in radiotherapy treatment planning~\cite{chen2022vrcontour} and surgery planning~\cite{gasques2021artemis}.
For acupuncture, a few existing commodity VR-based applications, such as AcuMap\footnote{AcuMap: \href{https://www.mai.ai/acumap}{https://www.mai.ai/acumap}.} attempted to utilize stereoscopic-rendered anatomy inside VR to assist novices in learning and practicing acupuncture workflow.
However, these systems often fall short to more experienced learners due to the inability to provide fine-grained visualizations of the medical structures and to adapt to a wide range of real-world human body variations such as age, gender, and disease-related changes. 
Medical imaging, like MRI and CT scans, plays a pivotal role in providing detailed internal views of the human body. They're especially effective in VR workflow due to their stereoscopic nature. For acupuncture training, it not only provides fine-grained details but also allows for the visualization of bodies with all kinds of variants, thus providing training material for customizing needling treatments individually. It also comes with the flexibility to highlight specific regions and anatomical structures with different visualization settings which can ultimately accommodate different levels of learners. 

In this paper, we present \textit{AcuVR}, an acupuncture training system that leverages the affordances of VR in addressing the limitations of traditional acupuncture training workflow. A key aspect of AcuVR is the integration and alignment of medical imaging (MRI/CT scans) and human anatomy models.
It mainly addresses the three shortages of common acupuncture training applications:
\textbf{(1)} Compared to the existing system which only provides one fake 3D modeling body for practice, AcuVR allows users to practice acupuncture based on real-world anatomical structure rendered by medical imaging data from real patients with various pathologies;
\textbf{(2)} Compared to the existing systems which only show needling points as 2D points lying on the surface skin, AcuVR shows 3d target points inside the body with the context of key anatomical structures. Thus, inserting a needle from surface skin in VR will effectively display the internal structures and depth it passes through.
\textbf{(3)} It allows for fine-tuning of visualization settings to pull out the regions and structures for specific purposes. These tuning methods are designed to accommodate different learner's proficiency in interpreting medical imaging. 

An iterative design approach was employed to incorporate feedback from four acupuncture experts and educators, and two students for three design iterations. Then we conducted a pilot study and interviewed eight practitioners and students, analyzing their feedback to assess AcuVR's effectiveness in enhancing acupuncture training workflows. Their insights highlight AcuVR's significant potential to overcome existing training challenges. Furthermore, our discussions reveal the adaptability of AcuVR's workflow to broader areas of medical training.
\section{Related Work}
\subsection{Acupuncture Training}
Contemporary acupuncture education programs integrate both classical theory and modern biomedical sciences~\cite{zhou2015dry}. These programs emphasize the importance of understanding human anatomy, physiology, and pathology, which can further enable students to work with other healthcare professionals~\cite{white2009western}. Although acupuncture is generally considered a safe practice, it can lead to various types of adverse events, including organ or tissue injuries, infections, local reactions, and body reactions (\eg dizziness~\cite{chan2017safety}). 
To ensure effective and safe treatments, expertise in Evidence-Based Medicine (EBM) is essential, as it integrates the best available scientific evidence with clinical expertise and patient values. In EBM education, cadaver-based approaches have long been recognized for their valuable contributions in providing insights into the shape, feel, and function of the human body~\cite{balta2017utility}. 
Moreover, studies have demonstrated that medical imaging of cadavers not only enhances students' comprehension of human anatomy but also improves their understanding of the relationship between acupuncture needles and anatomical structures. Works focusing on X-rays~\cite{leow2017use,kotze2012translucent}, CT scans~\cite{murakami2014integrated}, and Ultrasound~\cite{leow2016ultrasonography} have reported remarkable benefits, emphasizing the significance of medical imaging in acupuncture education. Another popular training tool in acupuncture education is the phantom acupoints model~\cite{jang2022trends}, which has demonstrated effectiveness in improving needle depth accuracy. 

Despite their broad utilization, these models have certain limitations. 
One notable drawback is the lack of visual feedback for evaluation during needle insertion~\cite{lee2015evaluation}. Visual cues are essential for students to observe and assess the accuracy of their needling techniques. 
This work presents the design and development of AcuVR, a training system that not only provides on-demand and intuitive acupuncture practice opportunities, but also enables continuous self-assessment by representing layers of human anatomy model, as well as linkage with medical images.

\subsection{Virtual Reality in Medical Education}

The affordability of VR technology has been adopted across many educational domains, revolutionizing the learning experience by providing immersive and interactive training opportunities~\cite{alfarsi2021general}. Notably, VR enhances performance and engagement and facilitates a more comprehensive understanding among learners compared to traditional methods~\cite{krokos2019virtual, radianti2020systematic}. 

In the medical education field, studies have shown that VR training outperforms traditional methods of learning anatomy, procedural skills, surgical procedures, communication skills, and clinical decision-making~\cite{turso2022virtual,liang2021analysis}. For example, a comparative study with 57 students investigated the educational effectiveness of a computer-generated 3D model that depicts the middle and inner ear and resulted in $18\%$ higher quiz scores~\cite{nicholson2006can}. 
In a more recent study, the use of VR training was found to enhance cadaveric temporal bone dissection performance by a significant $35\%$ when compared to traditional teaching methods~\cite{zhao2011can}. As VR enables simulation-based medical training that offers a safe experience for trainees to perform deliberate practice, it has increasingly been embraced in the realm of patient-based procedures~\cite{gunn2018use,ruthenbeck2015virtual}. This includes applications in dental and bone surgery~\cite{ayoub2019application}, minimally invasive surgery~\cite{guedes2019virtual}, and endoscopy training~\cite{khan2019virtual}, where the use of virtual reality technology has gained acceptance.

In terms of acupuncture training, traditional acupuncture training often relies on two-dimensional models, pictures, or auxiliary teaching aids to demonstrate the position of acupoints. This approach lacks details, intuitiveness, and accuracy, introducing significant cognitive load in the learning process~\cite{guan2022application}. To address these challenges, researchers combined medical anatomy education in VR and acupuncture knowledge to create a virtual three-dimensional space for teaching acupuncture~\cite{chen2019application}. This allows beginners to visualize the anatomical structure of acupoints, realistically simulate the acupuncture process, and provide a safe foundation to enhance clinical efficacy. For example, a teaching experiment at the Guangzhou University of Traditional Chinese Medicine utilized a virtual acupuncture teaching system to display anatomical details of key acupoints and meridian outlines (\ie strings connecting acupoints)~\cite{rao2020practical}. AcuMap~\cite{maiacumap} was the first commercialized acupuncture VR training system that incorporated anatomy, meridians, and acupoints on the standard 3D human model. While these systems have established benchmarks in anatomical precision, AcuVR distinguishes itself through its innovative integration of medical imaging and registration on standard anatomy in VR. Another trend in VR acupuncture research enables novice practitioners to localize acupoints on individuals with technologies including gesture recognition~\cite{zhang2017preliminary}, face alignment~\cite{zhang2021faceatlasar,zhang2022faceatlasar}, and full-body tracking~\cite{du2022mobile} to adapt to anatomical variations. 


\subsection{Development and Integration of Medical Imaging in VR}
Medical imaging plays a critical role in modern healthcare in both educational and non-educational contexts. It enables visualization and diagnosis of internal structures and conditions. Techniques such as Computed Tomography (CT) and Magnetic Resonance Imaging (MRI) produce detailed cross-sectional images of the body, aiding in the detection and characterization of diseases. These imaging modalities often generate data in the DICOM format, a widely used standard for storing, sharing, and exchanging medical images and related information among healthcare professionals and systems. 

Advanced visualization of medical imaging has become a significant area of research. The prevalence of 2D screens in DICOM viewers facilitates easy viewing of medical images. However, this approach poses challenges for clinical practitioners and students in extracting crucial 3D information, including anatomical structures, regions of interest, and spatial relationships. Today, there are many advanced DICOM viewer systems available that offer 3D volumetric visualization. These systems provide users with a more comprehensive view of imaging data, and they can be accessed on both desktop~\cite{prosurgicalweb,pieper20043d} and mobile platforms~\cite{osirixweb,postdicomweb}.

VR takes medical imaging visualization to the next level by creating immersive and interactive experiences. By integrating VR into medical imaging education, students can more effectively explore complex or subtle anatomy~\cite{maresky2019virtual} and engage in hands-on training simulations. At the same time, non-educational virtual reality in particular is being increasingly used for two main scenarios, such as realistic simulators and interactive medical data visualization stations~\cite{pires2021use}. A wide range of VR solutions already exist that integrate medical imaging with VR head-mounted displays. These solutions serve various purposes, including general visualization of patient-specific 3D data volumes and models~\cite{imagingreality,zhang2022directx,specto}. In addition, many applications are designed for specific medical workflows, such as surgical planning~\cite{surgicaltheater,zhang2021server}, simulation training~\cite{reznek2002virtual,gunn2018use}, and radiation oncology~\cite{immersiveOncology,chen2022vrcontour, Chen2022vrcontourneeds, Chen2022, Yarmand2024, Yarmand2023}, offering enhanced visualization and training experiences.

In terms of acupuncture, medical imaging not only aids in the accurate localization of acupoints, but also helps in understanding the mechanisms of acupuncture and its effects on the body~\cite{godson2019accuracy}. By incorporating medical imaging into acupuncture research and practice, researchers and practitioners can visualize and study the intricate relationships between acupoints, nerves, blood vessels, and other anatomical features~\cite{kim2012analysis, kim2014positioning,kim2015partially}. This integration promotes a more rigorous and scientific approach to acupuncture, strengthening its position as a viable therapeutic option within evidence-based medicine. For example, recent research at Harvard Medical School integrates meta-analysis and resting-state functional connectivity to identify potential locations of scalp acupuncture for the treatment of dementia~\cite{cao2020neuroimaging}.  

Despite the current effort towards integrating VR and medical imaging, research to date has primarily focused on the individual contributions of medical imaging in providing anatomical insights and VR in offering immersive, interactive experiences. However, their combined use, particularly within the context of acupuncture training, remains largely unexplored. AcuVR steps into this gap as a pioneering application, being the first VR system based on medical imaging specifically designed for acupuncture.
\section{Iterative Design}
\label{sec:iterative-design}

We employed a user-centered design approach and an in-depth engagement with domain experts to design the acupuncture workflow used in AcuVR. 

We recruited four acupuncture doctors and two students from a leading clinical and teaching institution in the United States that specializes in acupuncture training.

Over three sessions, we first familiarized the participating acupuncturists with the basic concepts of VR and its applications in medical domains (\eg Complete Anatomy~\cite{completeanatomy} and DICOMViewer for surgical planning~\cite{zhang2022directx}). We then presented existing computer-supported learning material for acupuncture (\eg AcuMap~\cite{maiacumap}). We then conducted hands-on sessions, in which the doctors and students tried out the functionality of the latest AcuVR version and reflected on their experience. Appendix table~\ref{fig:iterative-design} describes the three iteration steps, findings, and design considerations in more detail.

The participant comments were aggregated and grouped based on themes, following the affinity mapping protocol.

\subsection{Design Considerations}
Fig.~\ref{fig:iterative-design} shows detailed takeaways from each iteration, including the experts' general attitude towards this system and their suggestions on functionalities and interface design. Two key design considerations emerged from the iterative design:

\vspace{1em}
\noindent\textbf{\emph{Leveraging Layered 3D Standard Anatomy Models and Individual Medical Imaging Data Volumes for Acupuncture Training}}

\noindent Both 3D standard anatomy models and individual medical imaging scans are valuable materials for acupuncture training. 
The participants expressed that the 3D standard anatomy models are valuable learning sources especially for entry-level students to grasp basic concepts, yet individual medical scans can uniquely benefit higher-level learners to showcase complex points and techniques, as well as provide practice opportunities on real patient cases.  

Specifically, visualizing medical scans enables practitioners to examine anatomical structures and their unique variants, which help them analyze patient-specific variations and anomalies, and further ``tailor the treatments to be more specific'' [acupuncture doctor]. In the meanwhile, students and practitioners with limited radiology expertise may find interpreting medical images challenging. Intuitive and adaptable visualization controls for medical imaging, as well as facilitating cross-referencing between standard anatomy models and medical images can lower the learning barrier for novice students.


\vspace{1em}
\noindent\textbf{\emph{Modeling 3D Interactions Based on Medical Techniques}}

\noindent 
The iterative design process pointed to the incorporation of real-life methods of practitioners into the VR interaction techniques. Some participants lacked prior experience with VR, taking them up to 15 minutes to get used to the controllers for interacting with the user interface and objects in VR. Switching between different interaction methods~(e.g. switching between hands and controllers) also proved to be challenging for many participating doctors. To address these challenges, we iteratively fine-tuned the 3D interactions to be consistent with bare hands or controller hands (\ie holding controllers while performing like bare hands). Notably, participants demonstrated improved performance and preference when using bare hands, particularly during needling simulations, as it closely mimicked real-world scenarios and ``[feeling] so much more natural.''~[acupuncture doctor]. 
\section{System Design}
\label{sec:system-design}
Grounded on iterative design processes, we introduced three key features: (1) {\bf Medical imaging registration} that aligns the medical scans with 3D anatomy models for cross-referenced visualization; (2) {\bf Cross-dimensional visualization} that facilitates exploration with 2D and 3D visualization of the internal structures; and (3) {\bf Interactive needling simulation} that provides hands-on training experience.
With an example of needling on the head, we now demonstrate our design of these three key components.

\subsection{Users}

AcuVR is designed to aid individual learners at all skill levels in comprehending and honing their skills in acupuncture procedures: \textbf{Novices} are encouraged to start with 3D anatomy models to grasp fundamental structures, progressively moving to interpret medical imaging, enhancing their detailed structure understanding through direct comparison and overlay techniques; \textbf{Experienced learners} may delve into medical imaging for in-depth analysis, utilizing anatomy models for reference as needed, especially in complex diagnosis and customized planning. Both novice and experienced learners benefit from the interactive needling simulations. Experienced learners, in particular, can focus on identifying and targeting specific internal structures, with or without applying a needle.

\subsection{Medical Imaging Registration on Anatomy Models}

\begin{figure}
  \centering
    \includegraphics[width=\linewidth]{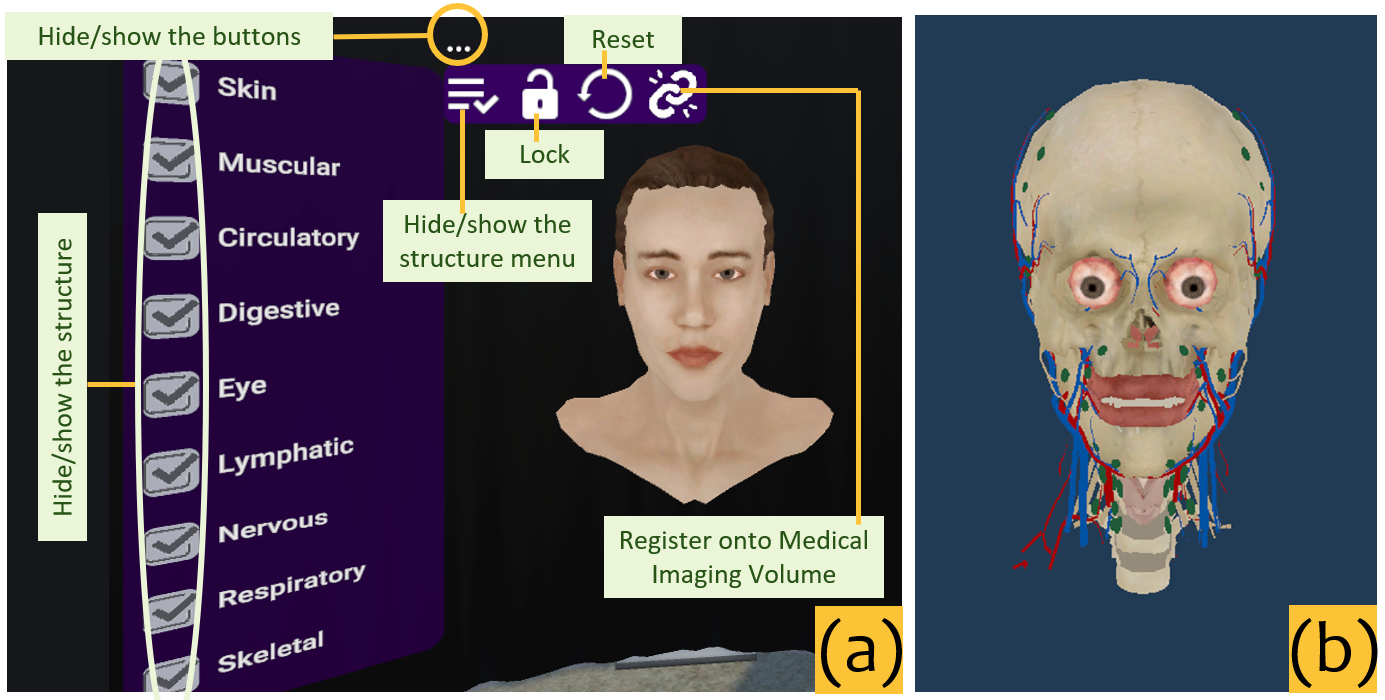}
    \vspace{-2em}
    \caption{Example of a 3D surface anatomy model: (a) a female head model with all anatomical structure layers, and (b) the same head model with selected structure layers.}
    \label{fig:standard}
\end{figure}

\vspace{0.5em}
\noindent\textbf{Layered Surface Anatomy Models} ---
The system incorporates surface anatomy models that are migrated from textbooks. These models present 3D layered anatomical structures, providing the basic cranial anatomy essential for acupuncture. For example, the layered head anatomy model shown in Fig.~\ref{fig:standard} enables users to toggle through structures ranging from the outer surface of skin to internal components such as bone, brain, organs, nerves, etc.

\vspace{0.5em}
\noindent\textbf{Medical Imaging Volume} ---
Surface models are limited to depicting the boundaries and contours of structures. In contrast, volume rendering reveals the intricate details of internal anatomical features and their depth. Therefore we introduce the volume rendering of medical images like CT and MRI to provide a more comprehensive understanding of internal anatomy.

\begin{figure}
  \centering
    \includegraphics[width=\linewidth]{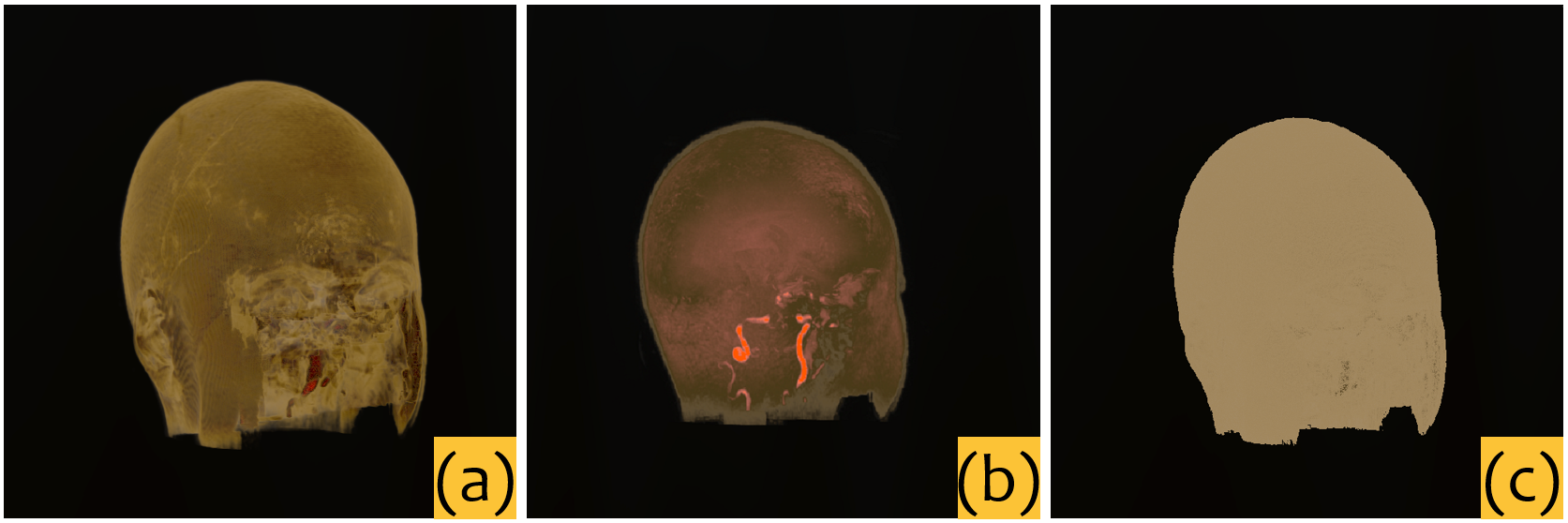}
    \vspace{-2em}
    \caption{A head MRI data volume visualization using: (a) direct volume rendering method to provide detailed spatial representations of soft tissue structures, (b) maximum intensity projection method to enhance high-intensity and high-contrast features, and (c) iso-surface volume rendering method to delineate boundaries between different tissue types.}
    \label{fig:volume-rendering}
\end{figure}

As shown in Fig.~\ref{fig:volume-rendering}, we provide three methods for medical volume visualization: Direct Volume Rendering (DVR), Maximum Intensity Projection (MIP), and Iso-Surface Volume Rendering to cover a spectrum of visualization capabilities: detailed tissue visualization, emphasizing high-valued structures, and creating clear boundaries. Among the three methods, DVR, which is the most commonly employed and serves as the default method, allows for the observation of complex anatomical details in a three-dimensional context. The implementation of direct volume rendering in our system leverages advanced algorithms~\cite{zhang2021server} to ensure real-time and high-quality visualization.

\begin{figure}
  \centering
    \includegraphics[width=\linewidth]{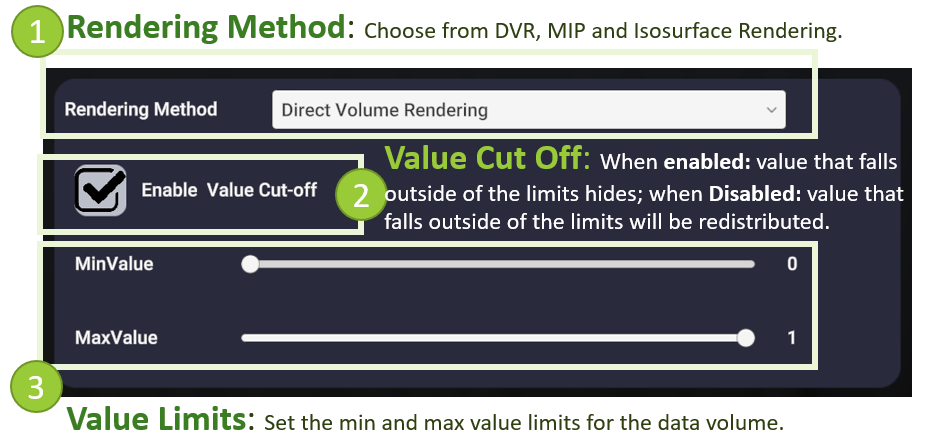}
    \vspace{-2em}
    \caption{Medical imaging visualization setting panel.}
    \label{fig:volume-setting}
\end{figure}

\begin{figure}
  \centering
    \includegraphics[width=\linewidth]{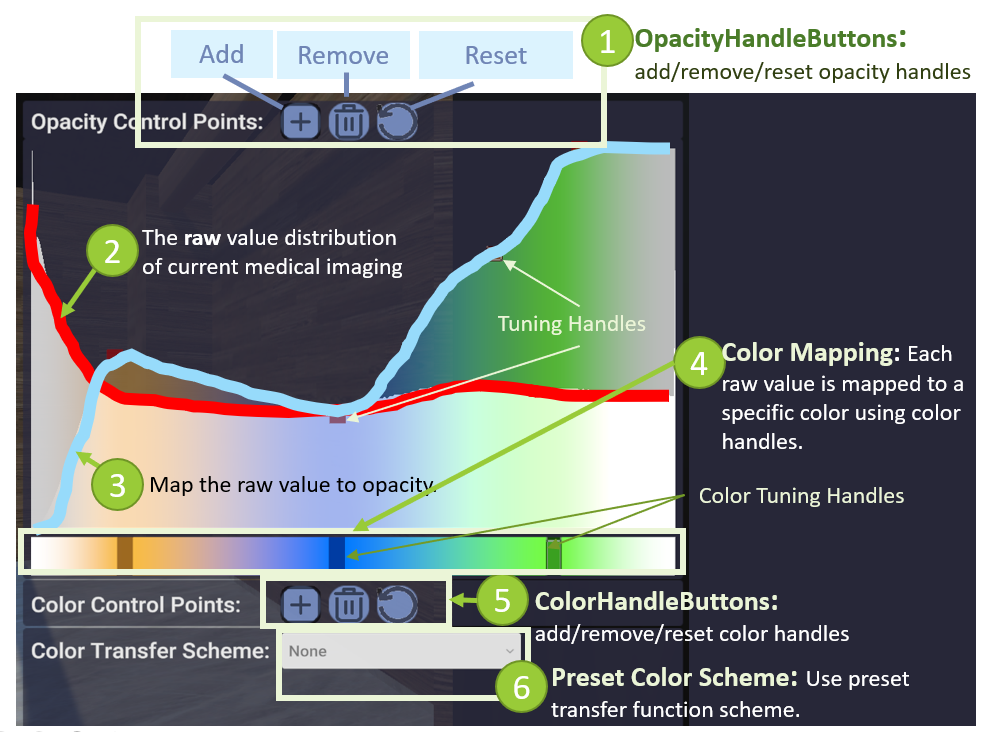}
    \vspace{-2em}
    \caption{1-Dimensional transfer function.}
    \label{fig:transferfunction1d}
\end{figure}

The best practice of using the DVR method involves the strategic assignment of varied opacity and color values to distinct tissues and materials, based on their voxel intensity values to distinguish between different anatomical structures effectively. Many existing medical training tools offer predefined setups for visualization, these preset color schemes, although simplifying the visualization process, restrict user flexibility. In contrast, our system provides complete control over visualization parameters, including contrast, intensity-opacity mapping, and color mapping~(Fig.~\ref{fig:volume-setting} and Fig.~\ref{fig:transferfunction1d}). This enables the application of customized transfer functions, where students can fine-tune the visualization to highlight their regions of interest (ROIs) precisely.

\begin{figure}
  \centering
    \includegraphics[width=\linewidth]{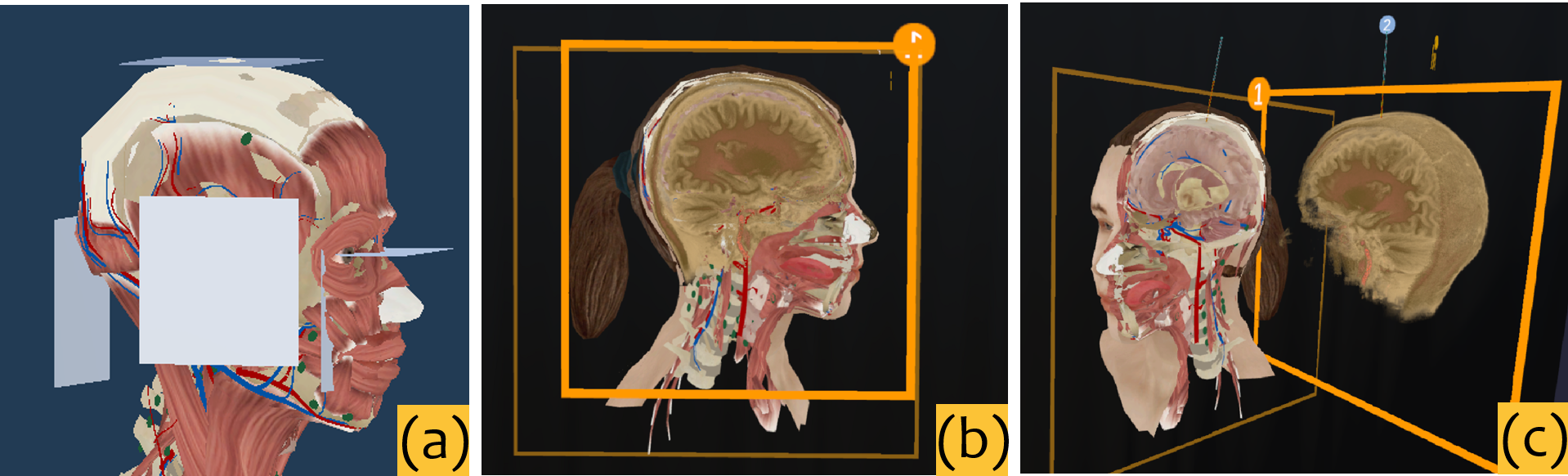}
    \vspace{-2em}
    \caption{Medical imaging - surface anatomy registration: (a) an example set of user-selected six anatomical landmarks, (b) overlapping registration representation, and (c) side-by-side placement.}
    \label{fig:registration}
\end{figure}

\vspace{0.5em}
\noindent\textbf{Registering Medical Imaging Volume On Surface Anatomy Models} ---
Students learn to discern the differences between observed medical imaging volumes and the standard anatomy model by comparing each medical imaging volume with its corresponding 3D surface anatomy. To enhance this learning process, we facilitate cross-referencing between the two materials by registering them together. Focusing on aligning the ROIs between the two entities, we employ a \textit{landmark-bounding-box alignment} method. Specifically, we enable users to:

\begin{enumerate}
\item Manually select six anatomical landmarks within each ROI to delineate its bounding box on both the standard model and target data volume. These landmarks typically consist of easily recognizable bony structures or organ landmarks that serve to define the boundaries of the ROIs. For instance, as illustrated in Fig.~\ref{fig:registration}a, the selected landmarks include the eyes (bottom/front), ears (left/right), and the top of the head skeleton (top) to align the patient's brain in medical imaging with a common educational model of a female head, enhancing the accuracy of anatomical comparisons and studies.
\item Align the two bounding boxes such that the position, rotation, and scale of the bounding boxes are the same, thereby registering the standard model to the data volume.
\end{enumerate}

Students can then examine the materials in either an overlapping mode~(Fig.~\ref{fig:registration}b) or a side-by-side~(Fig.~\ref{fig:registration}c) comparison mode.


\subsection{Cross-dimensional Visualizations for Supporting Explorations}
The integration of slicing planes (\ie ~cut-out and view planes) facilitates a multifaceted exploration of anatomical structures. This approach embodies the concept of cross-dimensional visualization by merging the clarity and detail of 2D imaging with the context and spatial awareness provided by 3D models.

\begin{figure}
  \centering
    \includegraphics[width=\linewidth]{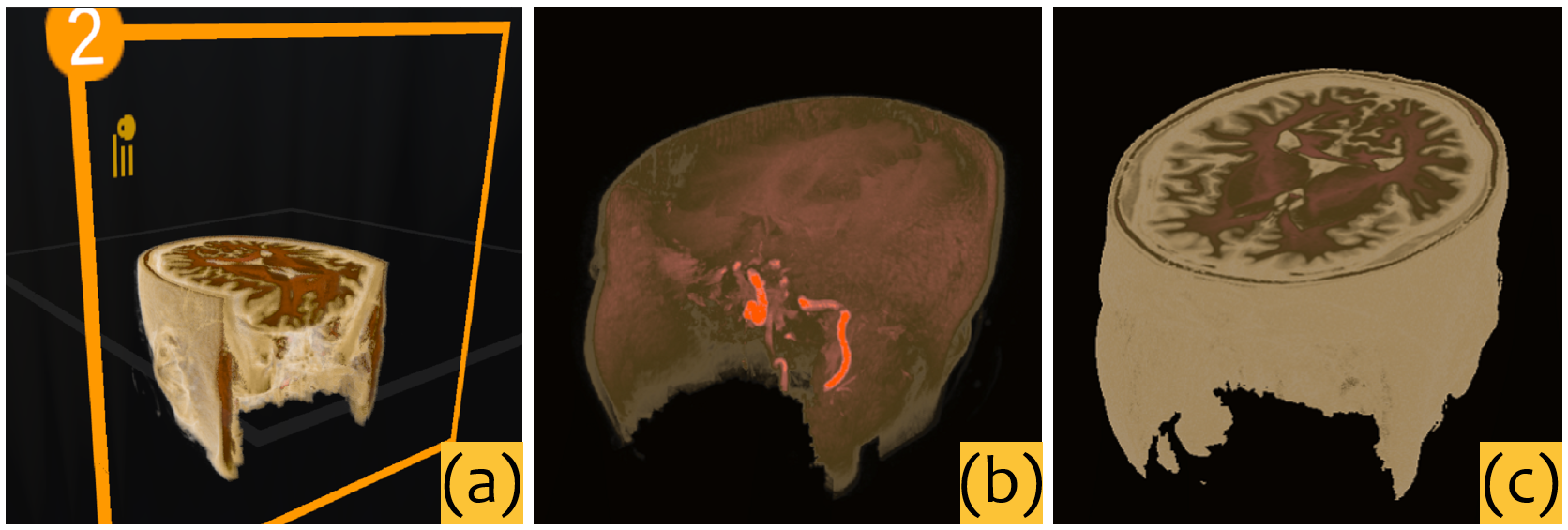}
    \vspace{-2em}
    \caption{Cut-out planes on the medical imaging volumes: (a) rendered with DVR method; (b) rendered with MIP method; and (c) rendered with isosurface method. Reveal cross-sectional 2D information by truncating the volume from cutting directions.}
    \label{fig:cutplane}
\end{figure}

\begin{figure}
  \centering
    \includegraphics[width=\linewidth]{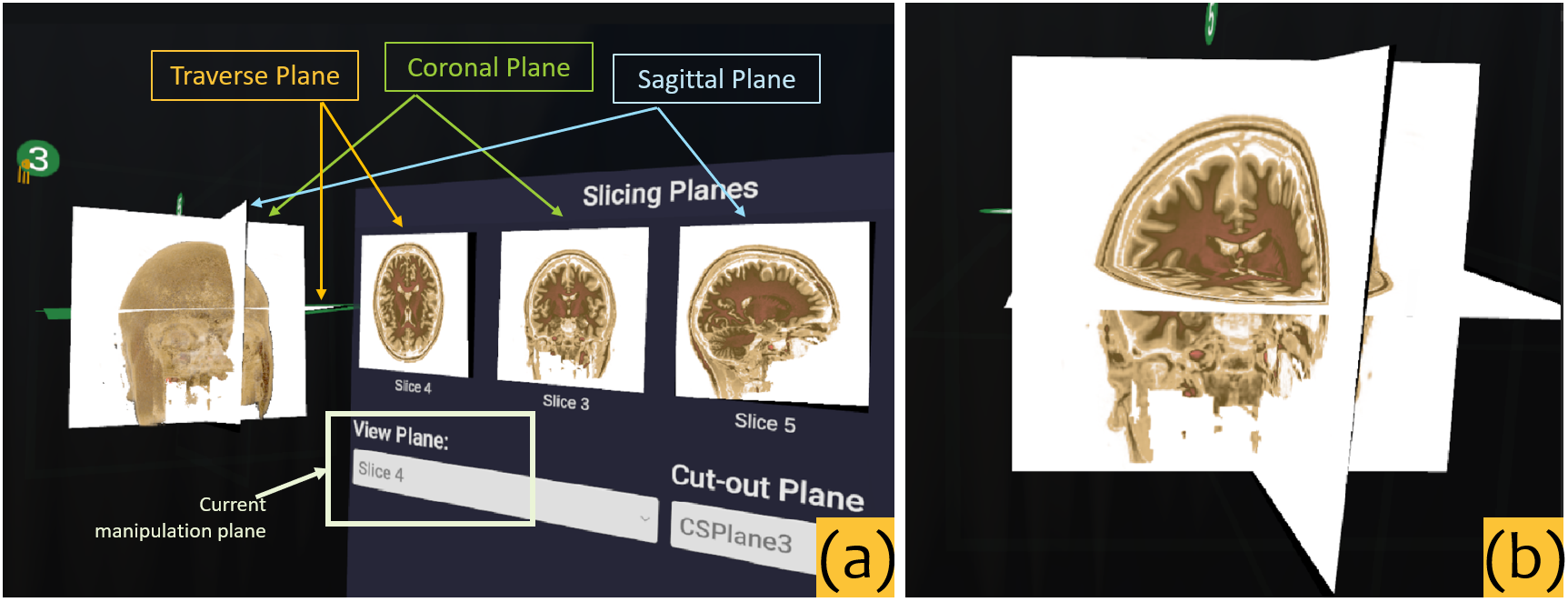}
    \vspace{-2em}
    \caption{View planes on the medical imaging volumes: (a) show three planes on the separate views displayed on the head-up UI panel; (b) visualization together with cut-out planes.}
    \label{fig:viewplane}
\end{figure}

\vspace{0.5em}
\noindent\textbf{Cut-out plane} truncates the volume at a specific location. It combines the depth and spatial relationships inherent in 3D models with the detailed focus of a 2D image plane. We provide six degrees of freedom for manipulation, where users can peel away layers of the model to expose hidden details, while still maintaining a sense of the overall anatomical context~(Fig.~\ref{fig:cutplane}). This capability is like looking through a window into the body, where both the surface being cut away and the internal structures revealed are visible simultaneously.

\vspace{0.5em}
\noindent\textbf{View plane} complements the cut-out plane by offering a direct, cross-sectional view of the medical imaging at a specific location, akin to traditional 2D CT/MRI techniques, but extends for six degrees of freedom too~(Fig.~\ref{fig:viewplane}a). It allows for an in-depth examination of the finer details within a single slice to support identifying specific features or pathologies that may require closer inspection.

\vspace{0.5em}
Both types of planes can be used together~(Fig.~\ref{fig:viewplane}b). The integration allows users to examine details across 2D and 3D information while utilizing both planes as a guide for needling operations to enhance practice accuracy.

\subsection{Interactive Needling Simulations}
The simulation features allow users to engage in comprehensive and realistic needling practices with self-assessment. 

\begin{figure}
  \centering
    \includegraphics[width=\linewidth]{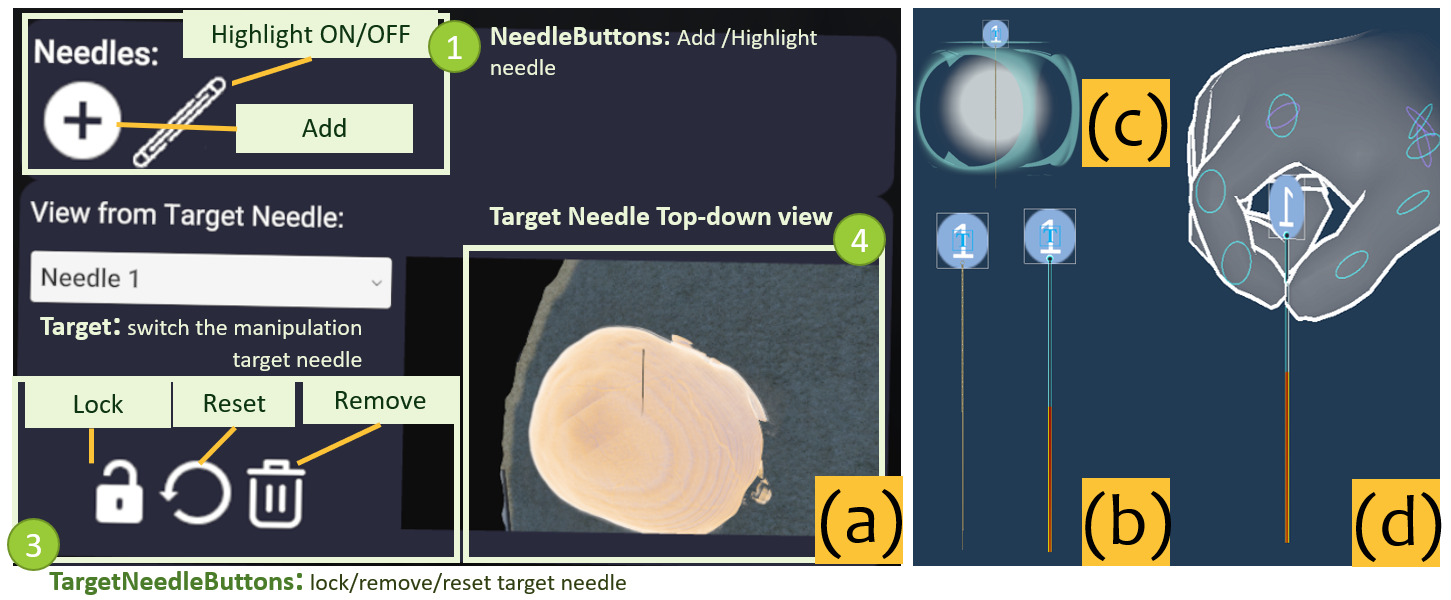}
    \vspace{-2em}
    \caption{AcuVR needling simulation: (a) head-up UI for needle selection and manipulation; (b) visualization of 40 mm acupuncture needles, with optional highlighting; (c) initiation of the needle; and (d) a synthesized hand holding the needle in scene}
    \label{fig:needles}
\end{figure}

\vspace{0.5em}
\noindent\textbf{Acupuncture needles} provided by AcuVR are designed to align the practice requirements on different parts of the body~(Fig.~\ref{fig:needles}b). The simulation replicates the tactile experience of needle insertion, requiring users to engage in the pinching motion akin to that used in actual acupuncture practices~(Fig.~\ref{fig:needles}d).

\begin{figure}
  \centering
    \includegraphics[width=\linewidth]{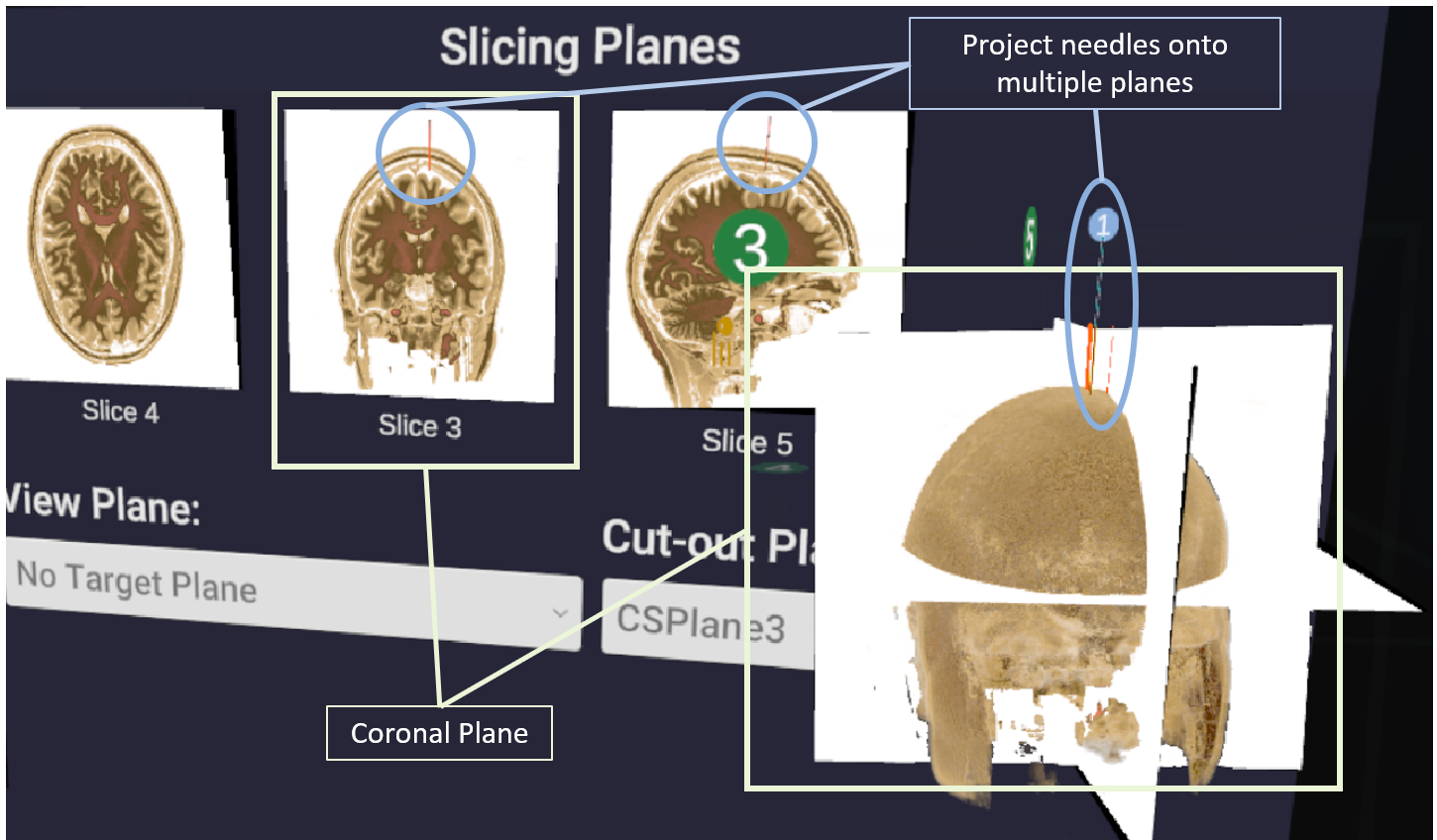}
    \vspace{-2em}
    \caption{Needle projection on view planes with position highlights, aiding users in verifying needle accuracy and placement.}
    \label{fig:needle-project}
\end{figure}

\vspace{0.5em}
\noindent\textbf{Detailed inspection} allows users to thoroughly examine needle placement once the needle has been inserted. We facilitate this inspection with a multi-view target needle visualizer~(Fig.~\ref{fig:needles}a) and view planes featuring needle projections. As shown in Fig.~\ref{fig:needle-project}, customized view planes display the projected path of the needles while highlighting the precise positions. This functionality is crucial for self-assessing the accuracy of needle insertions, ensuring that users can correct and refine their techniques in real-time 

\section{Study Methods}
we conducted an exploratory study that examined the following research questions:
\begin{itemize}
\item \textbf{RQ1} --- How does AcuVR, a medical imaging-based VR system, enhance acupuncture training workflow?
\item \textbf{RQ2} --- How does AcuVR complement the conventional learning curriculum of acupuncture?
\end{itemize}

\subsection{Participants}
To address these questions, we conducted a comprehensive study with eight acupuncture doctors and students~(Appendix Fig.~\ref{fig:demography}). All the participants in our study comprised licensed practitioners or senior students who possessed basic acupuncture knowledge and a fundamental understanding of acupuncture and anatomy. 

We specifically targeted this group of people to focus on challenging and risky tasks for AcuVR evaluation. Senior students, who already possess basic acupuncture experience, are at a stage where they are ready to tackle difficult points and techniques. 
 
In the pre-screening stage, participants were asked to complete a 5-point Likert scale questionnaire; they generally agreed ($87.5\%$ participants rate $\geq$ 3) that they were able to identify key structures from 3D anatomy, but disagreed ($87.5\%$ participants rate $\leq$ 3) that they have advanced proficiency in interpreting medical images. Any participant who has seen or tried the system~(Sec.~\ref{sec:iterative-design}) was excluded from this evaluation.

\subsection{Study Material}

\begin{figure*}
    \centering
    \includegraphics[width=\textwidth]{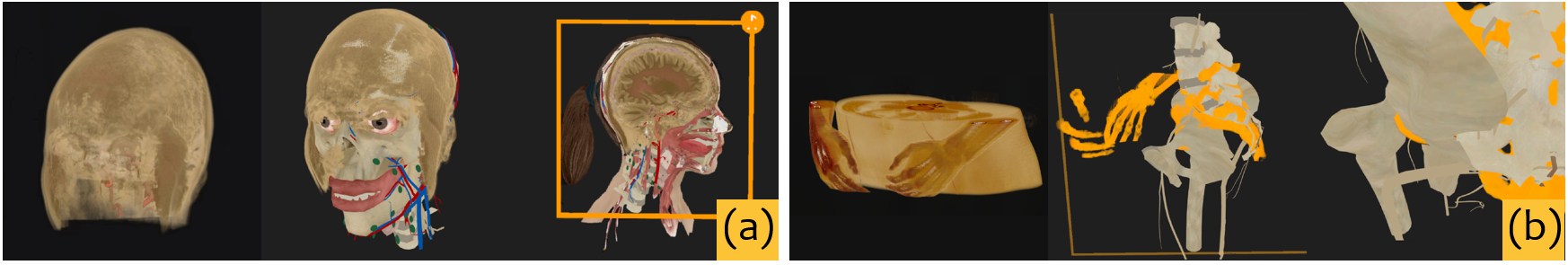}
    \caption{Two options for the pilot study: (a) a brain MRI and (b) a pelvis CT. In each case, a corresponding standard model is registered to the ROI.}
    \label{fig:study-mats}
\end{figure*}

With guidance from experienced doctors, we specifically chose two sample body parts for the user study: the head and the pelvis~(Fig.~\ref{fig:study-mats}). The DICOM images of the head are derived from the publicly available, anonymized NIH UPenn-GBM project~\cite{bakas2022university}, featuring MRI scans with pathology (glioblastoma). On the other hand, the pelvis images originate from the University of Iowa's Visible Human Project CT Datasets~\cite{iowact}, which consist of publicly accessible, anonymized CT scans of normal and healthy subjects.

In head acupuncture, needles are inserted shallowly into the scalp using \textit{needle threading}~\cite{hao2012review}, a technique that involves guiding a thin needle through specific scalp layers in precise patterns for targeted brain area stimulation. Understanding the structures and functions of various brain areas is essential for practitioners, hence the ROI covers the entire scalp~(Fig.~\ref{fig:study-mats}a).

In addition, the pelvis serves as an excellent example for demonstrating \textit{deep puncturing} needling techniques, which require precise insertion points located deep beneath the skin. These techniques can pose challenges even for experienced doctors due to variations and unique features in the pelvis. The emphasis here is on aligning the sacrum, particularly for deep puncturing targeting the sacral foramina~(Fig.~\ref{fig:study-mats}b).

\subsection{Before Study}
To ensure an efficient and focused study experience for the participants, we took into consideration their limited familiarity with VR technology. To expedite learning system functionalities, menus, and the user interface, We developed a 20-minute tutorial video that included video snippets showcasing each component described in Sec.~\ref{sec:system-design}. 
We used scalp acupuncture as an example to showcase each component's application in our system. The tutorial video concludes with a detailed needling demonstration on a pelvis acupoint, offering a scripted practice flow. This encapsulates the system's features, serving as a guide, particularly for VR novices. Participants were asked to watch this video before the study to learn the system's functionalities.

\subsection{Study Procedures}
We structured our study workflow as follows:

\vspace{0.5em}
\noindent
\textbf{Phase 1: Setup and Introduction}~(5 min) --- 
Participants completed a questionnaire on demographics and skills, and we customized the system setup based on their preferences. Unfamiliar participants received a quick system demo.

\vspace{0.5em}
\noindent
\textbf{Phase 2: Task-based Control Interaction}~(10--15 min) --- 
We directed participants to explore the VR system by performing basic actions with controllers or their hands. These tasks targeted the following:
\begin{itemize}
    \item the overall system, for instance, to reset and view directions
    \item control functionalities, including pinch, grab, and rotation
    \item image-based manipulation techniques, such as adjusting the transfer function menu and placing needles 
\end{itemize}

\vspace{0.5em}
\noindent\textbf{Phase 3: Cognitive Engagement}~(20 min) --- 
After completing the tutorial, participants were given the option to choose a specific body part~(head or pelvis) for free exploration in the system. We encouraged participants to engage in the \textit{think-aloud} protocol, where they freely explored the system while verbally sharing their thoughts and observations~\cite{van1994think}. Throughout the user study, we recorded videos to capture each participant's interactions for reference.

\subsection{Post-Study Interview}
After the participants completed the VR experience, a post-study questionnaire was distributed to gather their perception of the overall experience and cognitive Task Load Index~(TLX)~\cite{hart1988development} in six dimensions: frustration, effort, performance, temporal demand, physical demand, and mental demand.
Following that, we conducted semi-structured interviews to delve deeper into their questionnaire responses. Each interview lasted approximately 15 minutes and engaged in a discussion about the clinicians' opinions and experiences with guided questions. Our takeaways are later generated through inductive thematic analysis~\cite{braun2006using}. The guided questions for the overall experience evaluation dimension are:
\begin{itemize}
    \item When you rated your overall experience as X in the questionnaire, what was your main impression or reasoning behind your rating?
    \item In what circumstances do you anticipate using this system the most or least?
    \item What is the most significant difference you observed when comparing VR and traditional classroom methods?
    \item In what ways do you think VR could complement the traditional learning experience?
\end{itemize}
We also reviewed their answers to the TLX questions and asked the participants to articulate their reasoning behind some of the higher or lower-scored questions. We conducted a thematic analysis~\cite{braun2006using} using both deductive and inductive coding methods for qualitative evaluation. Initially, we transcribed and repeatedly reviewed the recorded audio clips, which allowed for the emergence of codes from the data. Throughout the analysis, we engaged in several discussions to review, iteratively refine these codes, and resolve any disagreements
\section{Results}
\subsection{Overall Usability and Task Load Insights}
On a scale of 1 to 5, all participants reported satisfaction with AcuVR: $50\%$ of participants "strongly agreed" and the remaining $50\%$ "agreed" that their experience with AcuVR was satisfying. When they were asked to evaluate the effectiveness of VR training for practicing needling skills in the target region, along with an enhanced comprehension of 3D acupoints and anatomy, the results showed that a considerable percentage of participants ($75\%$) "strongly agreed" or "agreed" that VR training facilitated more accurate needling of the targets ($37.5\%$ strongly agreed, $37.5\%$ agreed). Meanwhile, all participants unanimously expressed strong agreement ($100\%$) that VR training enhances their understanding of acupoints and anatomy after completing the training. 

\begin{figure}
  \centering
  \includegraphics[width=\linewidth]{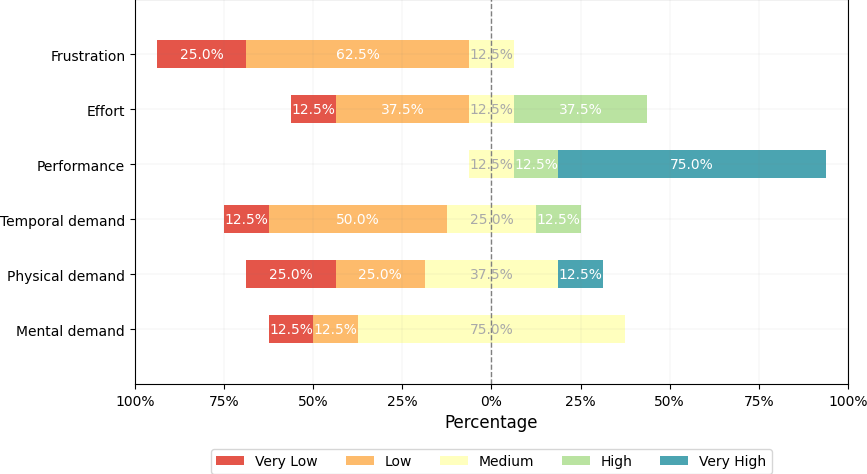}
  \caption{Task Load Index Responses: weighted workload categories. Ratings are depicted on a spectrum, with lower ratings presented on the left in warmer hues, and higher ratings on the right in cooler hues.}
  \label{fig:tle}
\end{figure}

Fig.~\ref{fig:tle} illustrates the distribution of responses to the Task Load Index, a key metric in assessing workload. It shows the distribution interpreted with the Weighted Workload category~(WWL)~\cite{colligan2015cognitive}, centered around the medium ratings. As shown, most participants evaluated their experience with the AcuVR system to exhibit high performance, yet low mental and physical demand, and minimal temporal demand and frustration.

Participants generally appreciated the well-designed integration of the standard model and medical imaging. Most positive comments about the system's usability revealed two main themes: ease of use and well-designed integration of functions. Most participants stated that AcuVR has a minimal learning curve. {\it ``There is not too steep a learning curve at all. I feel comfortable with the user interface''}~[P2] and {\it``I feel I pick it up easily''}~[P3]. They believed that most acupuncture doctors could easily use the system. Furthermore, P5 and P8 praised the needling experience, emphasizing its quality and smoothness; in particular, P8 stated, {\it ``Needling is my favorite feature. It was beautifully done, and the needling was very smooth"}~[P8].

On the other side, some participants mentioned the efforts in fine-tuning parameters to extract the regions of interest and enhancing visual distinction in structures, especially for those who are new to medical imaging inspection. For example, P7 expressed, {\it ``I didn't tune [the transfer functions] because I'm afraid I don't know how to do it without more instructions, it would still not be straightforward for me'}.

\subsection{Semantic Analysis}
Below, we present participants' perceptions of integrating AcuVR into their daily training:

\vspace{0.5em}
\noindent
\textbf{Advanced 3D Understanding Through AcuVR’s DICOM and Anatomy Integration} ---
The reliance on 2D textbooks and graphs is a major challenge in understanding the 3D contexts of the human body. AcuVR enhances participants' comprehension and spatial understanding of acupoints and anatomical structures. All participants appreciated the merits of exploring 3D structures in VR: {\it``it is visually helpful to understand in 3D, how far the needle is and where is it going into these different structures.''}~[P3]

The two different representations of anatomy (3D anatomy models and medical imaging volume) have distinct advantages. Traditional 3D anatomy model offers a simplified and {\it ``easiest way for me to use for [anatomical] structures, especially the ability to select the various systems (anatomical layers) in 3D''}~[P7].

On the other hand, the enhanced real-world medical imaging allows participants to {\it ``see real patients' bodies. Compared to [Existing training systems that] only provide a 3D anatomy model''}~[P2]. This feature not only enriches the material by {\it ``adding another dimension of the material''} especially for experienced learners, as highlighted by P7, but also the availability and the flexibility to adjust visualization settings in medical imaging are highly valued for their educational benefits. {\it ``By utilizing the contrast and gradient features and being able to compare different structures, students can learn about the actual structure and how it is modeled''}~[P7].
Despite the critical role of medical imaging in diagnosing and planning treatments for patients, it's still often overlooked in the current curriculum. As P3 highlighted, 
{\it ``only when someone who maybe has a Western background will be able to understand the patient's data better''}. Therefore, participants (P3, P5, P8) further noted that this training experience is particularly beneficial for improving their ability to interpret medical imaging to integrate knowledge of Western medical practices with traditional techniques.


The significance of integrating and aligning the two 3D representations was strongly emphasized, as they complement each other to create an efficient learning workflow. Participants noted this integration is particularly beneficial for identifying and understanding the locations of minuscule anatomical structures. This is crucial for practitioners to navigate these areas with greater precision, either targeted interventions or avoidance of specific regions (P2, P3, P4, P7). For instance, the ability to overlay nerves from the standard model onto medical scans was highlighted by P7, who found that this feature significantly aids in accurately locating nerves in relation to other vital structures within the scans. All participants predominantly engaged with the integration of medical imaging onto the 3D model, allocating their time significantly more to this feature over exploring medical imaging volumes or the anatomy model independently, at a ratio of 2:1:0.6 (see Fig.~\ref{fig:time-spent} for additional details). This usage underscores the potential advantages of such cross-reference between different educational materials. For example, {\it ``I like this link feature (registration) the best. Because I can’t recognize the vessel and lymph from the imaging, but I can use the model to hunt for them''}~[P6].

During the study, we noted that participants, especially novice students, employed the model-imaging registration to assess the accuracy of their needling techniques. Specifically, they would initially insert the needle through the skin surface without visual access to the underlying structures. Then, they removed each layer to verify whether the needle accurately avoided critical organs, bones, or nerves that should not be contacted. P1 is convinced that the training described above will equip students with the ability to {\it ``build an X-ray vision when a real person lying in front of you''}, which implies that practitioners will be able to intuitively visualize the skeleton and understand the spatial relationship between acupoints and bones beneath the surface of the skin. P4 believes such capability enables them to  {\it ``check [their] own biases''} while P7 believes it eases students' anxiety about potential mistakes

Experienced practitioners gain insights into both the practice and theory of acupuncture through AcuVR training. P7 noted that this training allows for a clearer visualization and explanation of how needling specific acupoints affects targeted muscles with visualization of anatomical layers. This insight not only deepens understanding but could contribute valuable perspectives to acupuncture theory, as highlighted by P1 and P7.

\vspace{0.5em}
\noindent
\textbf{Addressing the Main Obstacles of Acupuncture Students} ---
Participants pointed out three main obstacles to gaining safe hands-on needling practice in the current training workflow:
\begin{enumerate}[leftmargin=*]
    \item \emph{\underline{Time gap between theoretical learning and practicing}}: {\it ``We have to remember everything [from class] and wait until we are in clinics. This [system] helps us improve where we're at right now [along the class]''}~[IS1]. 
    \item \emph{\underline{Lack of opportunities to practice outside of class}}: Acupuncture practice requires supervision given the risky and specialized skills: {\it ``it's hard. You're not supposed to needle outside of class [without supervision]''}~[P3]. Yet, participants acknowledged their eagerness to practice: {\it ``they say, don't do it. But we all technically practice on your friends and family''}~[P2]. Many participants pointed out the on-demand benefits of AcuVR and compared it to a {\it ``virtual supervisor [that they can use to] practice on [their] own''}~[P6]. 
    \item \emph{\underline{Challenges of current lab-partner practice}}: Existing practice in class requires students needling on each. This is risky and can cause problems. For example, GB21 is a point on the shoulder where needling too deep can hurt the lungs. As P3 pointed out AcuVR would be really helpful for learning how to needle tricky areas like the lung safely. P5 further elaborated: {\it ``I think the key rationale is that you can visually see the context, the things that can be damaged by a needle.''}~[P5]. On the other hand, even when the needles are inserted correctly, {\it ``some people cannot be needled because they will have headaches, or vomit because of the practice''}~[P3]. Therefore, they often hesitate and lack the confidence to carry out the correct techniques, especially with challenging acupoints. For instance, in eye-acupuncture, students tend to be overly cautious, opting for {\it ``superficial needle insertion''} because the thought of needling a point so close to the eye is daunting (ID2). Therefore, they recognize AcuVR as {\it ``essential preview before needling my lab partner''}~[P6].
\end{enumerate}

\vspace{0.5em}
\noindent
\textbf{Enabling Customizable Training Workflow} --- When teaching students to treat patients with specific conditions, it's important to tailor acupuncture techniques based on the patient's disease. AcuVR, for example, uses brain tumor cases to show how MRI scans can guide the choice of acupuncture points and techniques. This approach helps illustrate how medical conditions affect treatment plans. Meanwhile, users aren't limited to preset examples; they can upload and study scans of different diseases in AcuVR, to enhance their understanding on how anatomy and illness influence acupuncture. P2 emphasized the importance of this personalized approach: {\it ``having access to the [the patient's] actual structure is essential. For example, if someone has undergone lower lumbar surgery and their vertebrae are fused, it becomes critical to avoid needling in the wrong place."} ID1 found our brain MRI example very valuable for practicing and applying their knowledge in cancer patient treatment:
\begin{quote}
    \emph{``I can see the range of the tumor here... I would treat locally and use the lines or points above the tumor to draw qi and blood to this area. This is a different practice from the theory of general scalp acupuncture, so here I can see where is the tumor boundary and then apply the needle.''}
\end{quote}

On the other side, Participants expressed benefits with respect to the standardization feature of AcuVR. This addresses the inaccuracies and uncertainties of practicing with lab partners whose unique physical traits or health issues could affect learning. P3 shared a personal insight: {\it ``Our lab partner could have all kinds of special issues like my partner who has scoliosis. I need to know how to needle a healthy one and set the standard with a healthy body."} 

The selected practice material on either healthy individuals or specific conditions provides a versatile foundation for students, which is challenging to achieve in current acupuncture classes. By integrating standard 3D modeling with public datasets showcasing standardized cases, AcuVR potentially offers a robust framework for enhancing teaching and setting benchmarks in acupuncture education.

\vspace{0.5em}
\noindent
\textbf{Augmenting the Palpation Learning with Visualization} ---
In acupuncture, palpation is essential for assessing the patient's body, locating acupuncture points, and detecting subtle changes~\cite{godson2019accuracy}. Practitioners palpate to find skeletal landmarks, assess conditions, and pinpoint acupuncture points, valuing tactile sensations rather than using modern technologies like ultrasound. Thus, linking palpation sensations to anatomical knowledge is crucial in learning.

Most participants indicated that AcuVR offers a comprehensive visualization of anatomy beneath the skin, and integrating palpation lectures with VR training could effectively establish connections between the visualizations and tactile sensations outside the skin.
%
Students~(P4, P5) expressed they want to have this system available in class along with the lecture for instructional purposes. {\it ``We'll be engaged if using it in the class. We can watch the professor [operate] and the professor can also connect to their model [in the real world]''}~[P5]. 
\section{Discussion and Future Work}

Overall, the findings of this study demonstrated the utility and versatility of AcuVR in enhancing acupuncture training workflow. It addressed the two research questions:

For \textbf{RQ1}, AcuVR enhances acupuncture training by providing a spatially accurate and immersive VR environment that integrates DICOM imaging data with standard anatomy models. This integration allows for a hands-on practice environment tailored to comprehend individual anatomical variations, which is crucial for all-level acupuncture training. The system visualizes needle placement under the skin in real-time while providing instant feedback. This ability helps learners refine their needling technique with self-assessment, and enhance learning through visualization. Such features lead to more precise acupoint knowledge and improved accuracy.

For \textbf{RQ2}, AcuVR potentially enriches the conventional acupuncture curriculum by facilitating self-training outside the classroom and integrating it into instructional processes during class. Its ease of use allows for seamless deployment, enabling students to explore acupoints and anatomical structures independently or through instructor-led demonstrations, thereby reinforcing understanding and facilitating a comprehensive approach to acupuncture practice.

\subsection{Broader Medical Training Context}
In discussing AcuVR's use in acupuncture, we uncover valuable insights for applying similar technologies in other medical training contexts, especially for those not specializing in medical imaging interpretation. AcuVR's novel approach to registering medical imaging volumes onto standard anatomy models has shown promise. This methodology has proven in pilot studies to assist individuals who are not adept at interpreting medical imaging by leveraging the synergy between realistic imaging data and standardized anatomical frameworks.

Our design of the workflow supports a learning curve, starting with pre-set visualizations for highlighting specific structures. This approach eases beginners into the complexity of medical imaging. Over time, enabling full control over visualization settings proves advantageous. It allows learners to independently identify and extract their own ROIs, tailoring their approach to each patient's anatomy and improving their medical imaging interpretation skills.

This progression from guided navigation to independent control enhances learners' understanding and diagnostic capabilities. AcuVR exemplifies how integrating innovative features into medical training can enhance practitioner skills, bridging the gap between textbook anatomy and real patient variations. This adaptive and personalized learning approach could serve as a model for expanding medical training tools beyond the field of acupuncture, emphasizing the development of nuanced medical imaging interpretation skills.

\subsection{Improving AcuVR}
Through observing participants utilizing AcuVR, we gained valuable insights and identified several areas for enhancement. 

\vspace{0.5em}
\noindent\textbf{Enhancing Palpation Simulation with Visual Cues} ---
An important aspect of evolving AcuVR is refining the palpation simulation using visual-based techniques. Simulating surface reactions to palpation through visual cues could offer students a more immersive learning experience. By mimicking the sense of touch visually, students can develop tactile skills without the need for physical contact. Some existing works incorporated haptic feedback devices, such as gloves or controllers to simulate tactile sensations. However, this option comes with extra setup, maintenance costs, and potential limitations on natural hand movements.

An alternative approach is the utilization of accurate real-time hand tracking. By employing advanced hand tracking technologies, this strategy can detect the precise placement of the user's hand on virtual content, such as the skin of the standard anatomy model or the 3D reconstructed CT/MRI scan. When the user places their hand on the virtual content, it can automatically highlight the underlying bony structure underneath, which closely mimics the palpation technique used in the real world.

\vspace{0.5em}
\noindent\textbf{Understand the Context in a More Dynamic Way} ---
Another aspect that is currently missing in the AcuVR system is the ability to understand the context of anatomy and acupoints more dynamically. While pre-scanned MRI/CT images cannot capture the natural movements of the body, there is potential to incorporate dynamic changes in the reconstructed models and standard 3D models. By introducing interactive elements that demonstrate how the anatomy and acupoints change during movement or specific conditions, learners can develop a more comprehensive understanding of the dynamic nature of acupuncture practices. This could involve showcasing variations in muscle tension and joint movement. 

\subsection{Limitations and Future Work}
One limitation of the user study was the variation in participants' familiarity with the VR training system. While efforts were made to inform all participants about the features and operation of the system before allowing them to explore freely, not all participants had watched the tutorial video before the study. This discrepancy in prior knowledge resulted in a difference in participants' initial understanding and comfort level with the system. This points to the importance of ensuring consistent training and preparation for all participants in future studies, to minimize any potential biases and ensure a level playing field in evaluating the system's effectiveness and usability.

In addition, the current phase of our research focuses primarily on the development and initial free-form tool exploration testing of AcuVR. In future work with the polished system, we intend to include a comparative study evaluating the effectiveness of AcuVR in educational settings compared to traditional acupuncture training methods and existing VR acupuncture training systems like AcuMap. With the collection and analysis of additional quantitative measures, including time of completion and accuracy of the final needling, we can further understand and demonstrate the system's potential to enhance learning outcomes in a comparable and qualitative manner.
\section{Conclusion}
    This paper presented AcuVR, a novel VR-based acupuncture training system that incorporates personalized medical imaging. In a pilot study (n=8), we explored the potential of AcuVR to enhance acupuncture training and improve the comprehension process, complementary to the traditional learning approaches. Feedback from participants indicated that AcuVR is easy to use, informative, and intuitive, showcasing its potential to benefit students in their advanced needling classes. By integrating medical imaging into the training system, AcuVR also offers a novel approach that enables students to gain a deeper understanding of acupuncture techniques and acupoint locations. The positive perception and user ratings point to a promising future for AcuVR as an invaluable tool for enhancing acupuncture education. As further advancements and refinements are made, AcuVR holds promising potential to revolutionize acupuncture training and contribute to the development of skilled acupuncture practitioners.

\balance
\bibliographystyle{ACM-Reference-Format}
\bibliography{references}

\clearpage
\appendix
\clearpage
\renewcommand{\thefigure}{A\arabic{figure}}
\setcounter{figure}{0}
\pagenumbering{roman}
\setcounter{page}{1}
\section{Three Iteration Steps}
\begin{figure*}
  \centering
  \includegraphics[width=\linewidth]{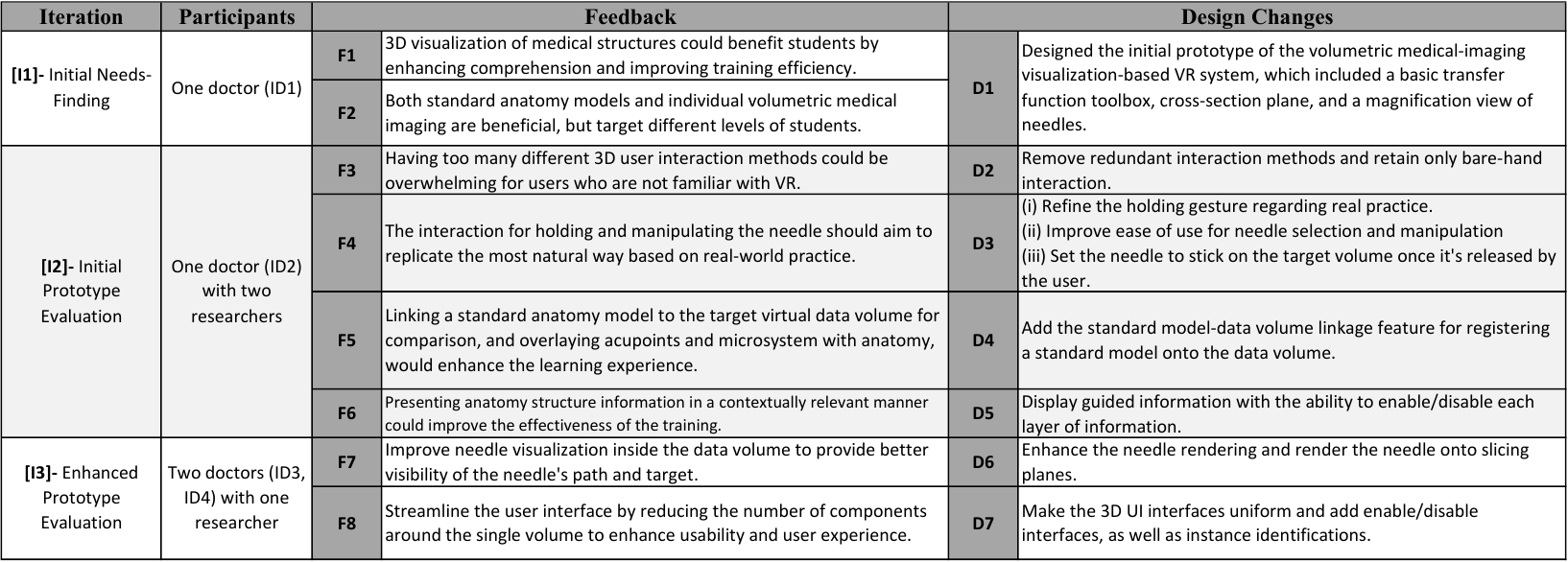}
  \vspace{-1.5em}
  \caption{Feedback insights and design consideration from the three iterative design. Iteration 1 established the basic framework and fundamental functionalities, while Iterations 2 and 3 focused on refining the details to enhance the usability and learnability of the prototype.}
  \label{fig:iterative-design}
  \vspace{-2em}
\end{figure*}

Fig.~\ref{fig:iterative-design} shows the three phases of the design iteration process and lessons learned.

\section{Participants Demography}

\begin{figure*}
  \centering
  \includegraphics[width=\linewidth]{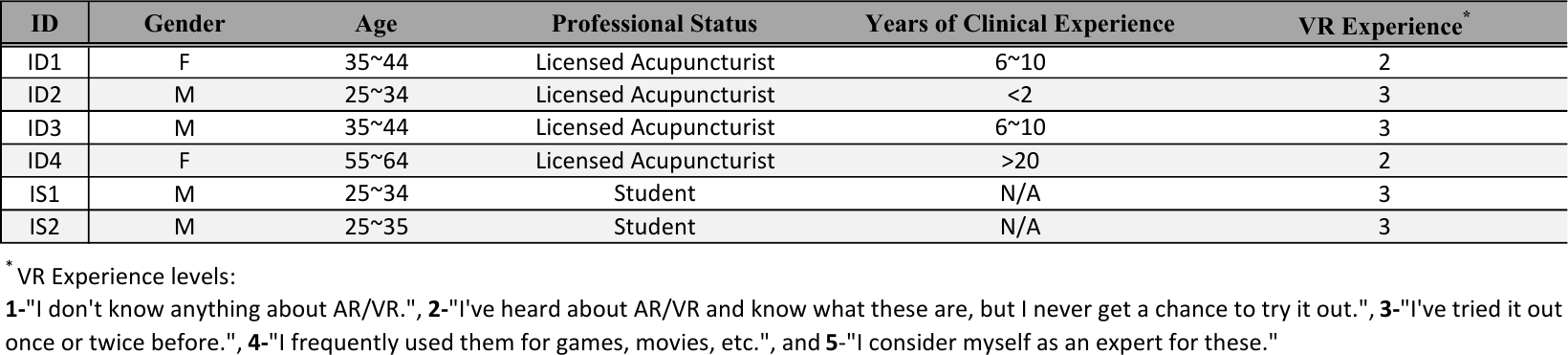}
  \caption{Demographics of the recruited participants during the iterative design process.}
  \label{fig:itr-participants}
\end{figure*}

\begin{figure*}[b]
  \centering
  \includegraphics[width=\linewidth]{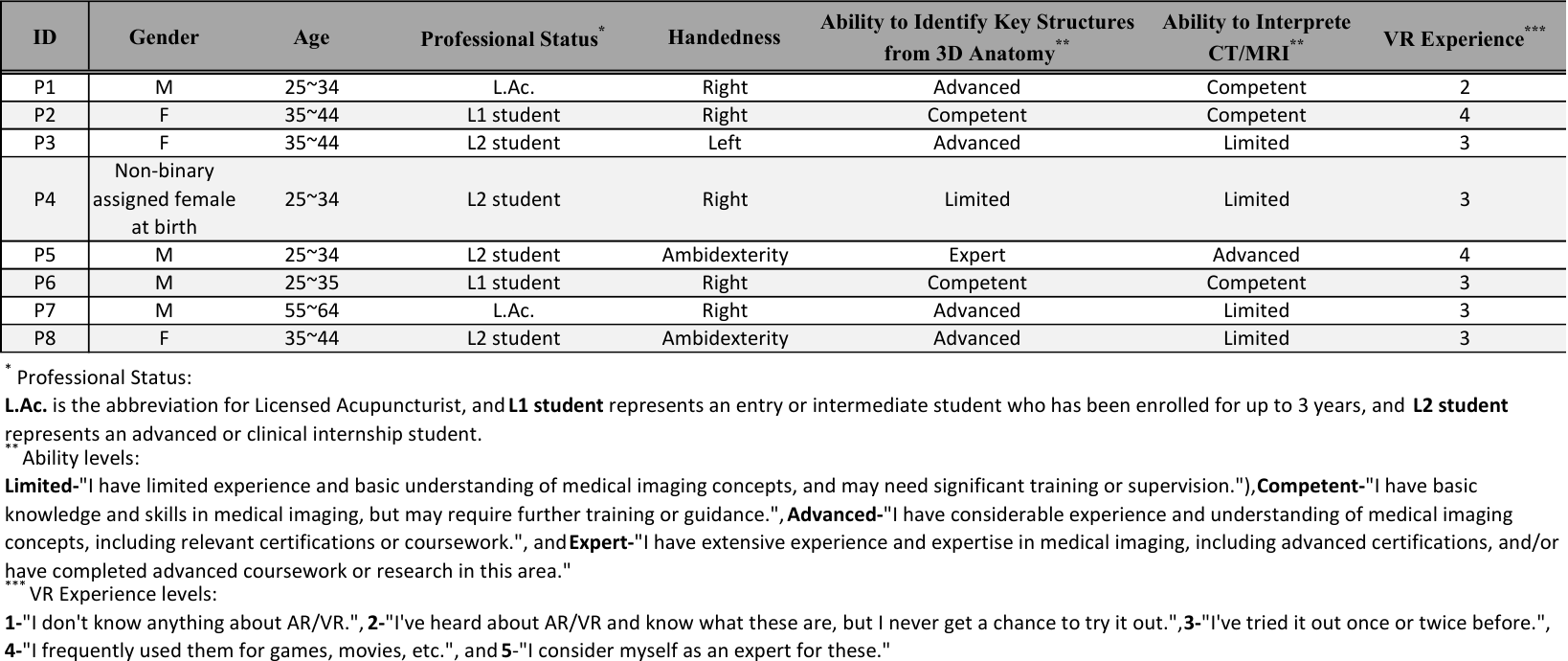}
  \caption{Demographics of the recruited participants.}
  \label{fig:demography}
\end{figure*}

Fig.~\ref{fig:itr-participants} and Fig.~\ref{fig:demography} show the demographics of the recruited participants during the iterative design process and user study.

\section{Task-based Control Interaction phase}
This section provides the components and tasks in the task-based control interaction phase.

\begin{table*}
\begin{tabularx}{\textwidth}{>{\hsize=0.15\hsize}X|>{\hsize=0.85\hsize}X}
\hline
\textbf{Component} & \textbf{Task} \\
\hline
System & Reset the system and view direction \\ \hline
Control 
& Look at your hands\\ \cline{2-2} 
& Pinch the volume to move and rotate it \\ \cline{2-2} 
& Grab the volume to move and rotate it  \\ \hline
UI & 
Press the button to hide the volume \\ \cline{2-2} 
& Bring up the Volume Property menu and use the slider to adjust the minimum value limit \\ \cline{2-2} 
& Bring up the Transfer function menu, add a new color widget, and adjust its position and color \\ \cline{2-2}
& Bring up the Slicing Plane menu, add a cut-out plane and a view plane, and manipulate their positions \\ \cline{2-2}
& Bring up the needle menu, add a new needle, adjust its position and angle, and place it onto the volume \\ \cline{2-2}
& Bring up the standard model menu, hide the skin, and link it to the volume \\ 
\hline
\end{tabularx}
\vspace{1em}
\caption{Components and tasks in the Task-based Control Interaction phase. These tasks were incorporated in the video tutorial, and later re-iterated during the study.}
\label{tab:study-task}
\end{table*}

\section{Study Results}

\begin{figure*}[b]
  \centering
  \includegraphics[width=\linewidth]{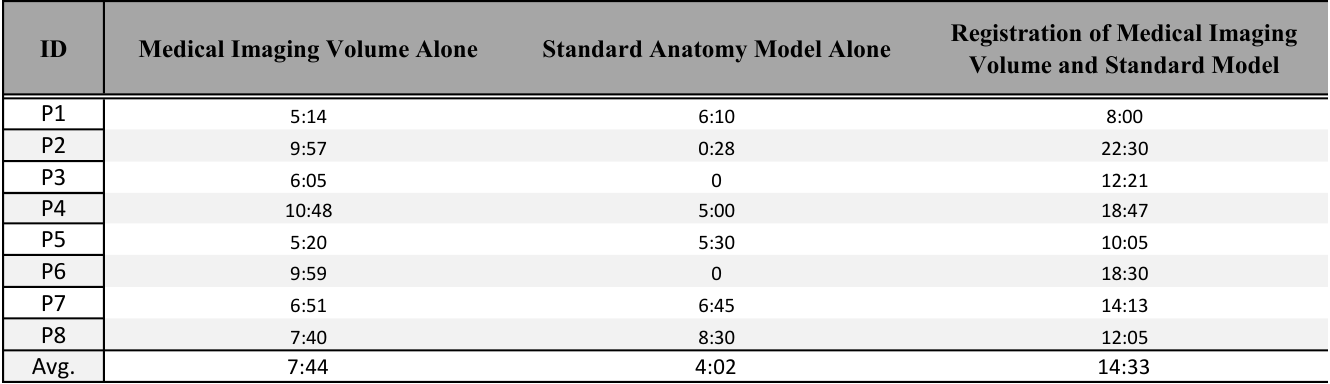}
  \caption{Time of each participant spent on the different types of 3D anatomy representation.}
  \label{fig:time-spent}
\end{figure*}

\section{Implementation Details}
\label{sec:system-methods}
In this section, we delve into the details of our system's development, complementing the information provided in Sec.~\ref{sec:system-design}.

We prototyped the system using Unity3D with Meta Quest 2 to leverage advanced hand-tracking capabilities. The system is optimized to run via Oculus AirLink~\cite{airlink} to guarantee high-quality volumetric rendering and smooth performance, allowing it to be synced seamlessly between the desktop and the headset for an immersive experience.

\subsection{Volumetric Rendering}
AcuVR contains three volume rendering methods: direct volume rendering~(DVR), Maximum Intensity Projection~(MIP), and Iso-Surface Volume Rendering~(Iso-surface) for different visualization purposes, as explained below:

\vspace{0.5em}
\noindent
\textbf{Direct Volume Rendering} --- The DVR method maps data volume intensity to visual attributes like color and opacity by taking into account the data values, transfer functions, and lighting conditions. The final color~($C(\mathbf{r})$) comes from integrating the volume rendering equation along each viewing ray, taking into account the data values~($\rho(\mathbf{r})$), transfer functions~($T(\mathbf{r})$), and lighting conditions~($L(\mathbf{r})$):
\begin{equation}
    C(\mathbf{r}) = \int_{\mathbf{r}_0}^{\mathbf{r}_f} T(\mathbf{r}) \cdot \rho(\mathbf{r}) \cdot L(\mathbf{r}) \, d\mathbf{r}
\end{equation}

\vspace{0.5em}
\noindent
\textbf{Maximum Intensity Projection} --- The MIP method selects the maximum intensity encountered along each ray, instead of accumulating the values. The core equation for MIP is:
\begin{equation}
C_{\text{{MIP}}} = \max(C_1, C_2, \ldots, C_n)
\end{equation}

\vspace{0.5em}
\noindent
\textbf{Iso-Surface Volume Rendering} --- The Iso-surface method extracts a specific threshold range and renders the corresponding surface:
\begin{equation}
f(\mathbf{r}) = 
\begin{cases}
    1, & \text{if } \rho(\mathbf{r}) > \text{threshold} \\
    0, & \text{otherwise}
\end{cases}
\vspace{0.5em}
\end{equation}

\subsection{Transfer Function}
Inspired by works of Chourasia et al.~\cite{chourasia2007data}, Lavik~\cite{lavik2020unity}, and Zhang et al.~\cite{zhang2021server}, we developed effective intensity transfer functions for high dynamic range scalar volume data. Our approach incorporates three key components:

\vspace{0.5em}
\noindent
\textbf{Contrast Enhancement} --- 
First, given a pair of contrast limits $(C_{min}, C_{max})$, the value of each voxel $C_x$ could either be cut off:
\begin{equation}
    C_{x'} = 
    \begin{cases}
    C_x, & \text{if } C_{min}\leq C_x \leq C_{max} \\
    0, & \text{otherwise}
\end{cases}
\end{equation}
or redistributed to enhance with a brightness factor $C_b$:
\begin{equation}
    \begin{split}
        C_{x'} = (C_x - C_{min}) / (C_{max} - C_{min}) + C_b,&\\
    (C_{min}\leq C_x \leq C_{max})&
    \end{split}
\end{equation}

\vspace{0.5em}
\noindent
\textbf{Linear Opacity and Color Mapping} --- 
Second, we utilize the intensity of the volume data to determine both its opacity and color, according to Lavik's work~\cite{lavik2020unity}. The x-axis represents density, while the y-axis signifies the corresponding opacity value. The underlying histogram aids in identifying the desired range for visualization, enabling users to add and adjust color and opacity handlers to refine the mapping. These interactions form a lookup table, allowing each intensity value to be mapped to a specific opacity and color value.

\vspace{0.5em}
\noindent
\textbf{Preset Color Scheme} ---
In addition to user-designed color transfer functions, we leverage the concept of fine spatial acuity, as human eyes are more sensitive to luminance gratings compared to chromatic gratings~\cite{mullen1985contrast}. Accordingly, we developed preset color schemes that ensure a uniform magnitude of incremental change in perceptual lightness. These carefully crafted color schemes contribute to enhanced detail perception, thereby improving the visualization of volumetric data.

\subsection{Needling Simulation}
\vspace{0.5em}
\noindent\textbf{Needle Projection} --- When users place a needle on the data volume, they can leverage the cut-out and view planes to aid in comprehending the underlying structures and see the needle's position. The view planes display the projections of the target needles, serving as visual cues to locate their positions~(Fig.~\ref{fig:needle-project}). This is achieved by projecting the tip and base of the needle onto all view planes. Specifically, for a view slicing plane $P$ with position $P_p$ and plane normal $P_n$, while the needle is defined by $V_1(x,y,z)$ and $V_2(x,y,z)$. We first find the closest point on the plane of $(V_1, V_2)$
\begin{equation}
   V_c = V - ((V - P_p) \cdot P_n) \cdot P_n
\end{equation}
and then project the point $V_c$ as:
\begin{equation}
    V_p =  V_c - (V_c\cdot P_n) \cdot P_n + V
\end{equation}

We then transform the projected point from the world space to the plane space and draw on the plane.

\section{Questionnaire}
This section provides the full pre-screening for the participants.

\begin{figure*}
  \centering
\includegraphics[height=20cm,keepaspectratio]{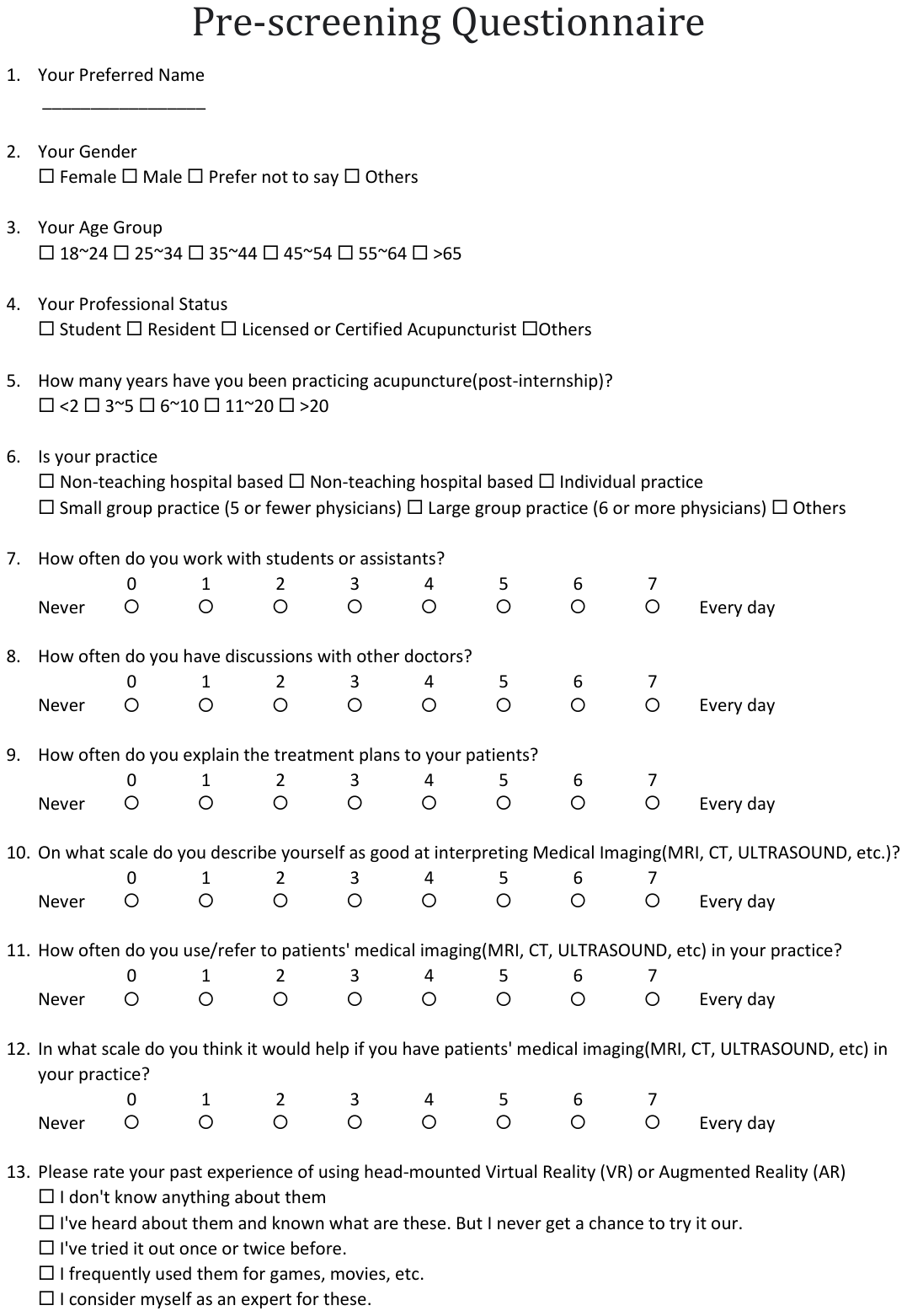}
  \caption{The 13-question pre-screening questionnaire.}
  \label{fig:pre-screening}
\end{figure*}
\clearpage


\end{document}